\documentclass{article}

\usepackage{arxiv}

\usepackage[utf8]{inputenc} 
\usepackage[T1]{fontenc}    
\usepackage{hyperref}       
\usepackage{url}            
\usepackage{booktabs}       
\usepackage{amsfonts}       
\usepackage{nicefrac}       
\usepackage{microtype}      
\usepackage{graphicx}
\usepackage{listings}
\usepackage{multirow}
\usepackage{url}

\title{Detecting Affective Flow States of Knowledge Workers Using Physiological Sensors}

\author{
  Matthew L. Lee \\
  FX Palo Alto Laboratory\\
  Palo Alto, CA 94304 \\
  \texttt{matthew.l.lee@acm.org} \\
}

\hypersetup{
pdftitle={Detecting Affective Flow States of Knowledge Workers Using Physiological Sensors},
pdfsubject={},
pdfauthor={Matthew L. Lee},
pdfkeywords={affective computing, physiological sensing, flow, workplace, productivity, emotion},
}

\begin{document}
\maketitle

\begin{abstract}
Flow-like experiences at work are important for productivity and worker well-being. However, it is difficult to objectively detect when workers are experiencing flow in their work. In this paper, we investigate how to predict a worker's focus state based on physiological signals. We conducted a lab study to collect physiological data from knowledge workers experienced different levels of flow while performing work tasks. We used the nine characteristics of flow to design tasks that would induce different focus states. A manipulation check using the Flow Short Scale verified that participants experienced three distinct flow states, one overly challenging non-flow state, and two types of flow states, balanced flow, and automatic flow. 
We built machine learning classifiers that can distinguish between non-flow and flow states with 0.889 average AUC and rest states from working states with 0.98 average AUC. The results show that physiological sensing can detect focused flow states of knowledge workers and can enable ways to for individuals and organizations to improve both productivity and worker satisfaction.
\end{abstract}

\keywords{affective computing \and physiological sensing \and flow \and workplace \and productivity \and emotion}

\section{Introduction}
Countless articles offer lifehacks to "beat distractions and stay focused" at work. Indeed, the workplace is increasingly littered with focus-robbing distractions from the environment, other people, digital notifications, and self-interruptions. Workers are particularly susceptible to distractions when their work tasks are too easy (leading to boredom) or too difficult (leading to stress). Workers tend to be most productive and efficient when they can work in a focused, flow-like mental state on a meaningful and reasonably challenging task. Indeed, higher job performance is linked with the positive psychology concept of a flow state, the subjective experience of engaging in just-manageable challenges that seamlessly unfold from moment to moment, particularly in conscientious individuals \cite{demerouti2006job}. Positive emotional states like flow can broaden a worker's thought-action repertoires, enabling better task performance \cite{fredrickson2004broaden}, greater job satisfaction, and higher job commitment \cite{lefevre1988flow}. In contrast, negative emotional states, such as stress and burnout due to overly challenging work, account for nearly one million people per day missing work, which results in \$50  to \$300 billion annually for U.S. employers \cite{stress_workplace_2011}. Likewise, boredom resulting from lowly engaged workers leads to depressive complaints, job dissatisfaction, absenteeism, and higher turnover rates for organizations \cite{kass2001watching} \cite{van2014boredom}. 

Surveys show that workers typically spend less than 10\% of their work day in a focused flow-like state \cite{cranston2013}, indicating an opportunity to unlock untapped productivity from existing workers while increasing satisfaction and well-being. Workers and employers can make adjustments to the work environment, job tasks, and the worker's skills to facilitate flow states. However, workers and organizations are often not aware of when workers are in or out of a focused flow-like state and lack an understanding of the contexts in which workers are able to be most productive. Therefore, dynamically detecting and measuring a worker's level of focus on their work tasks would enable both the employee and the organization to adjust work practices, including organizational interventions (such as additional training or task-reassignment) or individually-targeted technological interventions (like blocking distractions, goal setting and reminding) that can help workers enter into and maintain a focused flow-like state for a greater percentage of their workday.  

However, it is difficult to measure the level of focus workers have when performing work tasks. Traditional methods such as experience sampling can disrupt a worker's attention and task performance. Post-hoc diary studies are difficult because subjective feelings of focus or flow states are transient and difficult to recall and report. Previous work in measuring flow states in game players \cite{peifer2014relation} and drivers \cite{tozman2015understanding} have shown some relationships between certain physiological signals and flow-like states. In our work, we investigate and characterize the focus states of information workers as they perform work-related tasks.

In this paper, we conducted a lab study to collect physiological data from information workers experienced different levels of flow while performing work tasks. We used the nine characteristics of flow to design tasks that would induce different focus states. A manipulation check using the Flow Short Scale verified that participants experienced three distinct flow states, one overly challenging non-flow state, and two types of flow states, balanced flow, and automatic flow. 
We described machine learning classifiers that can distinguish between non-flow and flow states with 0.889 AUC and rest states from working states with 0.98 AUC, averaged across participants. Finally, we will discuss how a measure of flow can be useful in future applications in the workplace to help workers enter into and maintain a sense of flow in their work.
\section{Background} \label{Background}
\subsection{Flow in the Workplace} \label{Flow in the Workplace}
The concept of flow was first introduced by Csikszentmihalyi \cite{nakamura2014concept} as an optimal experience of intrinsically motivated activities and has been described as "the subjective experience of engaging in just-manageable challenges by tackling a series of goals, continuously processing feedback about progress, and adjusting action based on the feedback." \cite{nakamura2014concept} Studied in various domains such as sports, writing, scientific creativity, and driving, flow experiences have been identified to have nine different phenomenological characteristics: 
\begin{enumerate}
\item \textit{Challenge/Skill Balance} - the task difficulty compared with a person's ability to do a task, with optimal experiences having a somewhat challenging task paired with a somewhat high level of ability, requiring a small stretch of skill. 
\item \textit{Clear Goals} - knowing clearly what needs to be done proximally and how to do it. 
\item \textit{Automaticity of Actions} - the feeling that actions performed feel automatic, with the self merging with actions.
\item \textit{Unambiguous Feedback} - immediate and clear feedback from the task itself that reinforces that worker is getting closer to the goal.
\item \textit{Sense of Control} - having the required and desired means to deal with whatever comes up to perform the task.  
\item \textit{High Concentration} - the subjective feeling of focused, almost effortless attention to the task.
\item \textit{Transformation of Time} - the perception that time is moving faster or slower than normal. 
\item \textit{Loss of Self-consciousness} - not feeling concerned or focused on the task as a way to represent who one is.
\item \textit{Autotelic Experience} - performing the task itself is intrinsically enjoyable or meaningful.
\end{enumerate}

The first characteristic, Challenge/Skill Balance, has a central role in flow experiences \cite{massimini_systematic_1988}. When the perceived challenge of a task far exceeds the person's skill, the task will be "too hard" resulting in anxiety, stress, and excessive cognitive load. Attention can drift from the task and the actions at hand to the self and one's shortcomings, making it even more difficult to engage with the task. At the other extreme, When the perceived challenge of the task is far below the person's skill, the person experiences boredom. To adapt to the under-stimulation, the user's attention tends to drift away from the task to more engaging stimuli. Flow experiences result only when the perceived challenges balances with (or only slightly exceeds in an attainable way) the person's abilities. Csikszentmihalyi et al. conducted a experience sampling study and found that flow experiences occur more often in high-challenge, high-skill situations \cite{csikszentmihalyi1989optimal}.  

In the specific context of work, flow has been described as "a short-term peak experience at work that is characterized by absorption (total immersion), work enjoyment and intrinsic work motivation" \cite{bakker2005flow}, with autonomy and performance feedback acting as antecedents to flow experiences. Fullagar et al. \cite{fullagar2009flow} in a study using hierarchical linear modeling found that 74\% of the variance in flow ratings at work was due to situational rather than dispositional factors, indicating that opportunities to facilitate more flow experiences may lie in modifying tasks or the work environment rather than changing the individual alone. Flow experiences at work lead to well-being outcomes by satisfying the basic psychological needs for competence and autonomy \cite{ilies2017flow}, with the resulting positive states increasing cognitive resources and acting as a buffer against future negative experiences \cite{fredrickson2004broaden}. Overall, flow experiences at work have positive benefits in terms of productivity and well-being for employees and employers. 

We note that flow is a subjective experience with varying degrees rather than simply a binary state. In this paper, when we refer to being in flow, flow states, or flow experiences, we mean an experience with a relatively high degree of flow characteristics compared to ordinary work experiences. We acknowledge that workers may not experience the highest degree of flow with work tasks, as they might in their hobbies, athletic pursuits, or leisure activities where their intrinsic motivation and challenge/skill balance may be higher \cite{engeser2016fluctuation}.

\subsection{Physiological Correlates of Flow} \label{Physiological Correlates of Flow}
Developing unobtrusive ways to detect when people are in or not in flow experiences is the first step in understanding how to specifically adjust the worker's tasks or work environment to facilitate more flow experiences, productivity, and well-being for workers and their employers.

One way to unobtrusively detect whether or not a worker is experiencing flow is to measure a proxy--the physiological parameters that correlate to the flow state. In prior research, Gaggliloi et al. \cite{gaggioli2013psychophysiological} have found that flow is associated with moderate arousal as measured by elevated heart rate, decrease in R-R interval mean, and increased heart rate variability (LF/HF ratio). Interestingly, these observations are typically more associated with situations of stress rather than relaxed states, indicating flow state is its own unique state of positive (rather than negative) arousal. Similar observations were found in Peifer et al.'s psychophysiological model of flow \cite{peifer2014relation}, demonstrating an inverted-U function of physiological arousal and flow experience, with low and high cardiovascular arousal indicating a state of relaxation or a state of stress respectively and a moderate arousal associated with flow. Note that these observations were made for flow states during daily activities rather than for work tasks. Flow state at work may differ from other activities, and further research is required to show whether these observations hold true in the workplace. 

In the domain of driving, heart rate variability was found to be correlated with flow states. Tozman et al. conducted a study \cite{tozman2015understanding} that manipulated the demands of the driving task between too easy, too difficult, and balanced (inducing flow) similar to our study approach, HF-HRV and LF-HRV in the balanced condition were found to be moderately lower than in the too easy condition and significantly higher than in the too difficult condition. In contrast to Peifer et al. \cite{peifer2014relation}, a linear relation between LF-HRV and flow was found, instead of an inverted-U shape. The exact interpretations of why certain HRV metrics reflect flow states are still being debated. In our work, we focus specifically on flow during tasks designed to reflect day-to-day work activities, and build machine learning models to identify flow states at work.


In studies of music playing, self-reported flow was related to decreased interbeat intervals, increased cardiac output, increased respiratory rate, increased respiratory depth, and activation of facial muscles, in particularly, decreased activity in the corrugator supercilii used in frowning and increased activity in the zygomaticus major used in smiling \cite{de2010psychophysiology}. Further, increased flow was related to increased LF/HF ratio, total power of of HRV. This suggests that during a cognitively demanding task, an increased activation of the sympathetic branch of the automatic nervous system combined with deeper breathing and smiling might be an indicator for flow states. Expert musicians during a challenging task (playing a new musical piece) exhibited an increase in HF-HRV, longer interbeat intervals, and higher respiratory rate, indicating that experts were more able to recruit their parasympthetic nervous system to moderate their arousal to deal with the novel challenge \cite{ullen201010}. 

In summary, measures of arousal and parasympathetic/sympathetic activity seem to be important for characterizing flow states across multiple domains. However, the exact relationships with flow found in prior studies do not always align. Consistent with the phenomenological experience of flow, moderately high (rather than very low or high) levels of arousal are associated with flow states. However, in some cases parasympathetic activity alone as measured by time-domain and frequency-domain HRV metrics was correlated with flow and in other cases both sympathetic and parasympathetic activity were associated with flow. 

\subsection{Cognitive Load and Stress} \label{Cognitive Load and Stress}
A construct related to flow states is cognitive load, the amount of mental effort exerted when performing a task. Prior work has looked at quantifying cognitive load using physiological sensors in multiple domains. In psychometric studies using tasks such as mental arithmetic and dual n-back \cite{jaeggi2008improving}, low frequency components of HRV \cite{rowe1998heart} as well as time-domain components of HRV \cite{cinaz2013monitoring} have been found to correlate with levels of cognitive load and stress. In studies of driving, average heart rate increased, interbeat intervals decreased, and pupil diameters increased during increase cognitive load \cite{dong2011driver}. Pupil size was also positively correlated with task difficulty in studies of mental workload and interruptions on work tasks \cite{bailey2008understanding} and was detectable using video-based eye tracking \cite{klingner2010measuring} and even under varying luminance conditions \cite{wang2013indexing}. From studies of educational games, flow experience has been found to be negatively correlated with intrinsic cognitive load (i.e., the perceived difficulty of the material or in our case, the work task) and positively correlated with germane cognitive load (i.e., constructing and automating schemas for material) \cite{chang2017game}. Thus, we should expect that work tasks that demand a high cognitive load would be low in flow experiences. In our work, we attempt to identify how flow experiences differ across office work tasks of various difficulty levels.

\subsection{Measuring Flow in the Workplace}
In the specific domain of flow at work, prior works have used different methods to automatically detect and quantify flow experiences at work. Ara et al. \cite{ara2009predicting} attempted to predict flow state in daily work through continuous sensing of motion rhythm. After a month-long study in which office workers wore sensor badges with 3-axis accelerometers while self-reporting flow ratings, they found that motion rhythm around 2-3Hz was moderately correlated with the richness of flow during work, though the sign of the correlation differed across individuals. It is not clear why certain rhythm frequencies were associated with flow states in the above study. Further, motion rhythms may be measuring a distal proxy of flow experiences. Instead, in our approach, we detect and quantify flow using a worker's physiological signals which has the potential to be more direct and also less specific to the types of motions in the workplace in the above study.

Other works have explored the connection of particular digital work and platforms to flow states. For example, in a survey of Amazon Mechanical Turk users by Bucher et al. \cite{bucher2017flow}, performing microwork can result in flow-like states of immersion, particularly if the worker is given enough autonomy, experiences a good challenge/skill balance, and receives significant feedback. The flow states enjoyed by performing clearly defined tasks might even compensate for the relatively low monetary incentives that mTurkers receive. In Mauri et al.'s study \cite{mauri2011facebook} of Facebook users, researchers measured different physiological signals including skin conductance, interbeat intervals, EEG, EMG of facial muscles, respiratory activity, and pupil dilation under three conditions--using Facebook, performing a stress-inducing Stroop test, and during relaxation. Facebook use was associated with a mix of physiological characteristics from both the stress and relax conditions, having similar skin conductance to stress, the interbeat interval of relaxation, and the facial muscle activation normally associated with positive valence. The researchers characterize this combination of positive valence and relative high arousal as a flow-like state that helps users engage successfully with Facebook. A study by Müller et al. \cite{muller2015stuck} that sensed the physiological signals of software developers and found that developers experience a range of emotions at work and positive valence is correlated with perceived progress. Physiological sensors could predict the affect (positive or negative valence) which can be used to predict whether developers were stuck or making progress. The findings were focused more towards positive valence and higher arousal rather than predicting flow. In our work, we focus more directly on measuring the physiological response in flow and non-flow states and building a model to predict flow states rather than emotional valence.

While not directly measuring flow states, prior work has looked at related constructs at work such as feelings of focus or boredom. A study by Mark et al. \cite{mark2014bored} that used experience sampling and computer use monitoring revealed that certain work activities were associated with certain attentional states. For example, email was associated with feeling focused whereas surfing the internet and window switching was associated with feeling bored. Using the dual dimensions of engagement and challenge, they segmented attentional states into focused, rote, bored, and frustrated. The construct of high engagement (comprising both 'focused' and 'rote' attentional states) seems to be similar to the type of flow we consider in our study. 

Considering this review of the literature on flow experiences and the workplace, there still does not exist a robust way to detect flow experience in the workplace, despite the obvious benefits to the worker and organization to have a better way to detect, quantify, and optimize for flow experiences, leading to better productivity and well-being. In our research, we build on prior work in other domains such as psychology, gaming, and driving, with our approach to unobtrusively sense the physiological correlates of flow and non-flow states with workplace-appropriate sensing to predict flow and non-flow states in workers.




\section{Research Questions}
In our work, we aim to answer the following research questions:
\begin{enumerate}
\item How can we induce flow states in the work place setting by manipulating work tasks? 
\item How well can we classify a worker's focus state based on physiological parameters captured by unobtrusive sensors?
\end{enumerate}
\section{Methodology} \label{Methodology}
To answer our research questions, we ran a data collection experiment to gather data on how workers' physiological signals change in different focus states, verified that workers experienced different focus states, 
and then built and evaluated a machine learning model that classifies a worker's focus state based on physiological features. First, we describe the data collection experimental procedure. 

To collect a valid dataset of physiological data labeled with their corresponding focus state, we either had to identify a worker's focus state or induce a focus state and then collect the worker's physiological responses during those states. We initially considered a field study in which users would wear sensors and self-label chunks of their work day with their perceived focus state. However, a field study would provide relatively sparse data because workers may forget to label chunks of their day (or be unwilling even when prompted with experience sampling). A worker's focus states may change quickly as the worker self-interrupts, switches tasks, or spends energy, and we were not initially confident that an experience sampling technique would capture labels at a fine enough resolution without being annoying. Based on these considerations, we decided to collect physiological data within a laboratory setting in which we induce a focus state in workers by manipulating the choice and parameters of typical work tasks. In the following sections, we describe how we designed tasks to induce different focus states in workers and what physiological data we collected using sensors. 


\subsection{Inducing flow and non-flow states with tasks}
To design tasks that will induce different focus states that can be considered flow or non-flow, we selected tasks and manipulated their characteristics according to the the nine characteristics of flow states (see \ref{Flow in the Workplace}).

According to most theories of flow states, the first characteristic, Challenge/Skill Balance, is often one of the strongest factors that determines whether a worker is in a bored, anxious, or flow-like state. Prior work from psychology used working memory tasks such as dual n-back as challenging tasks to induce cognitive load \cite{jaeggi2008improving}. However, this type of abstract tasks does not represent meaningful work tasks typically performed for longer periods. With our study participants being researchers and graduate students in mind, we selected three work tasks that spanned the space of challenge/skill balance: editing a spreadsheet (low challenge, high skill, resulting in 'boredom'), reading patent specifications and answering questions (high challenge, low skill, resulting in 'anxiety'), and summarizing a research paper (moderate-high challenge, high skill, resulting in a flow-like state). For each of these tasks, we designed specific elements to intentionally facilitate or block the characteristics of flow. 

For the boredom-inducing "too easy" task, the Spreadsheet task, participants were told to monotonously edit a series of 12 spreadsheets files to remove non-alphabetic characters from a list of email address. Each spreadsheet file had a single column of 30-80 email addresses. Participants had to manually edit addresses such as "john.doe12@gmail.com" to "johndoe@gmail.com" using the keyboard and mouse. We manipulated (i.e., reduced) the sense of control by not allowing the participant to use search and replace or write a script, which is what all participants would have normally done. We tried to reduce the automaticity by splitting the email addresses into different files, requiring participants to save, close, and open files to resume editing. Feedback was also meant to be ambiguous because participants did not know how many addresses were in each file, so we expected them not to know whether they were going to accomplish the task or not. As a "paperwork" task, it was designed to be neither meaningful nor enjoyable. 

For the anxiety-inducing "too difficult" task, the Patent task, participants were given 12 patent specifications (PDFs) and a list of questions to answer based on the patent contents. The researchers in our study were not experts in reading the difficult legal language in patent specifications, and we selected a series of technical patents covering drilling and extracting techniques used in hydraulic fracking, an unfamiliar topic that many generally associate with unpleasant environmental impacts. To limit the sense of control, we prevented participants from using the search function or using the internet to find the answers. To reduce the clarity of goals, the questions were purposely worded ambiguously so as not to provide clues to which patent document or section to look in. Neither feedback on their progress or the correctness of their answers was given on the task, even as they typed in their answers in a MS Word document. We also expected the automaticity of actions to be low because participants had to constantly make the conscious decision of continuing reading the current patent or open a new unread patent to find the answer. 

For the flow-inducing task, the Research Task, participants were told to choose a research paper that they would be interested in reading and to summarize it in a few powerpoint slides. Participants were researchers so this task was something that required at least a moderate-high skill in understanding a paper from their specific domain, resulting in high balance between challenge and skill. The task goals were specified fairly clearly to be a summary of the technical and scientific contributions of the paper as well as the relevance to the participant's own work. For a high sense of control, participants were given free reign to use whatever tools they wanted, including searching the internet or taking notes. 
Reading and summarizing a paper is a fairly well-rehearsed task for our participants, so we expected it to result in a high automaticity of actions (reading, taking notes, writing title slides, bullet points, etc), feedback was fairly unambiguous because they knew how long the paper was and how much content has been summarized so far. Marshalling attentional resources for a high level of concentration should be easy for reading an interesting paper, compared to the other tasks. We also expect this Research task to be somewhat intrinsically rewarding or enjoyable because participants selected the paper themselves, often a paper they would read for work anyway. 

\begin{figure} 
\centering
  \includegraphics[width=6in]{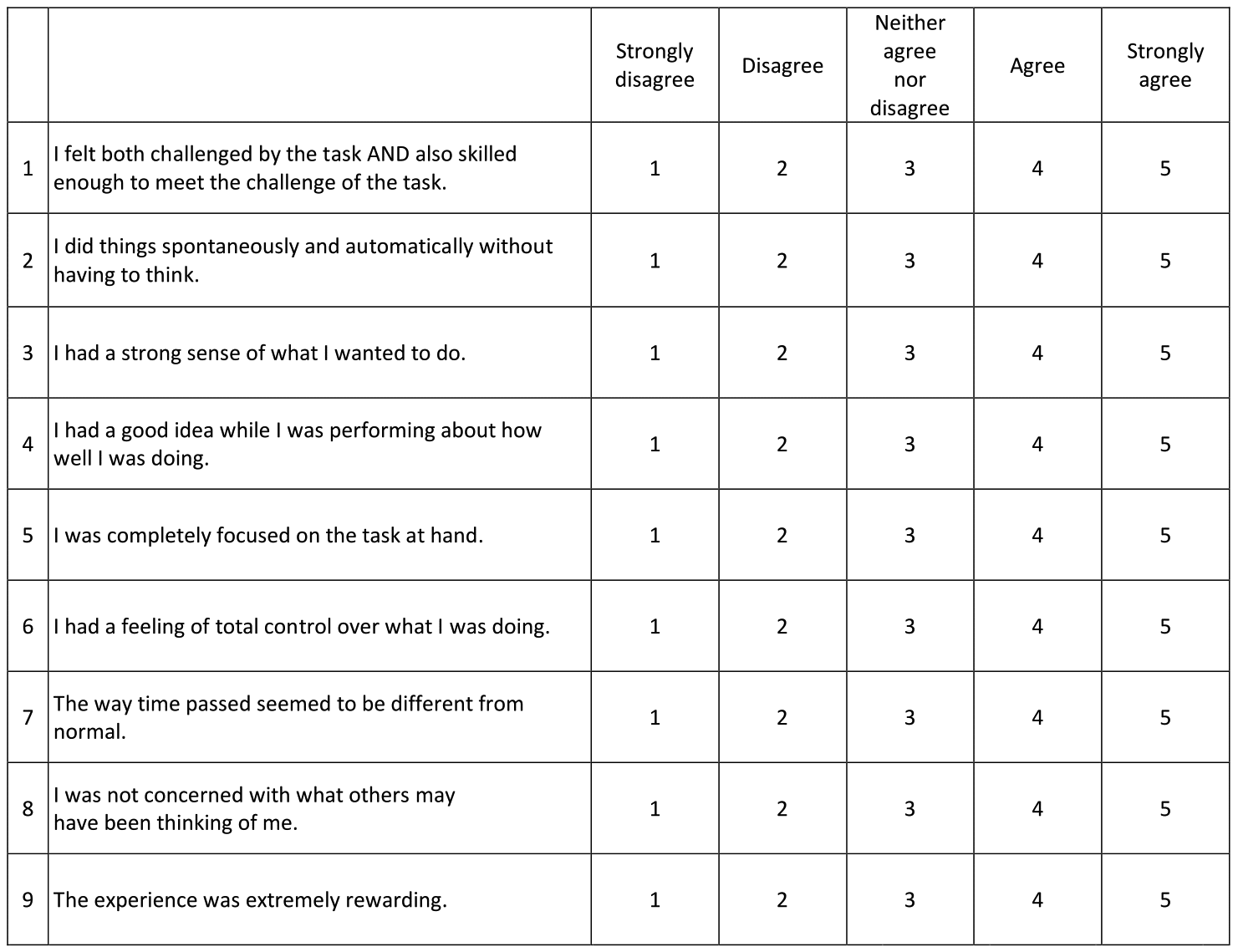}
  \caption{The Flow short scale (FSS) questionnaire used for the manipulation check}
    \label{fig:fss}
\end{figure}

The tasks were pilot tested with workers not included in the main study and tweaked to be as realistic and focus-state inducing as possible. To verify how successfully the designed tasks were able to induce different characteristics of flow-like states during the actual data collection, we had participants complete the Flow Short Scale (FSS) \cite{vollmeyer_motivational_2006} (Figure \ref{fig:fss}) after performing each task to rate how they felt along each characteristic of flow during each task. The FSS consisted of 9 statements, each corresponding to a characteristic of flow, and participants had to rate on a scale from 1 (Strong Disagree) to 5 (Strongly Agree). 
The results of the manipulation check can be found in \ref{Manipulation Check}. 

\subsection{Study Participants}
Study participants were selected to be representative information workers who regularly work on the computer. In our case, we selected workers with a research background including industry research scientists and graduate students due to (admittedly) easier access to them but more importantly so that flow-inducing task (e.g., reading a research paper) would be at the right level of challenge. Whereas we would have preferred to include other types of workers, it was difficult to design a single task that would be at the right challenge/skill balance for different types of workers. The main inclusion criteria were at least one year of regularly reading research papers and basic familiarity with editing Excel spreadsheets. Participants who have had extensive experience reading patent specifications were excluded, though we did not encounter any in our recruitment. 

Participants were recruited from an industrial research lab and from a nearby university. Participants received a USD50 Amazon or Starbucks gift card as a thank you for their participation. Using emails to distribution lists and snowball sampling, we recruited a total of 17 participants, and excluded 5 participants. Two of the exclusions were due to failing to meeting the inclusion criteria (which we only found out once we brought them into the lab). One participant was excluded due to language issues where reading a research paper in her non-native English was unexpectedly more difficult than expected. The remaining two exclusions were due to a system malfunction resulting in a loss of sensor data. At the end, we successfully collected data from 12 participants (9 males, 3 females), with an age range of 22 to 49. 

\subsection{Experimental Procedure}
\begin{figure}
\centering
  \includegraphics[width=1\columnwidth]{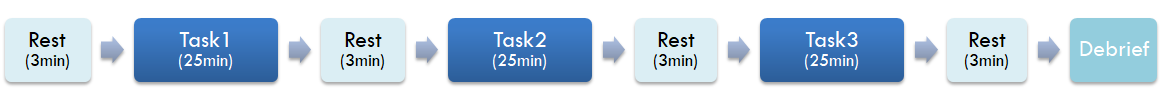}
  \caption{Experimental protocol consisted of a series of tasks lasting 25 minutes and rest periods. The order of tasks were fully counterbalanced among participants. }
    \label{fig:experimental-prototol}
\end{figure}

The study session (Figure \ref{fig:experimental-prototol}) comprised an introduction, three 25-minute work tasks with rest periods in between, and a post-interview with a researcher. The order of tasks was fully counterbalanced so that each of the six permutations of task orders were covered by two different participants. Overall, the study session lasted between 1h45m and 2h.  

In the introduction, participants were told about the purpose of the study, what is being tracked, and what to expect in the study. After participants gave their consent to participate in the study, the researcher explained how to wear the sensors (described in the next section \ref{Physiological Sensors and Measures}), and checked that they were worn and calibrated properly. Participants adjusted the workstation so they were seated comfortably and could see the display clearly. 

The data collection began with a rest period in which the participant sat silently and watched a slide show of static nature photos on the computer for 3 minutes. After the initial rest period, the participant was given the instructions for the first task on a piece of paper which they read to themselves. The researcher pointed out where the appropriate files are on the computer and clarified any questions the participants might have had. Participants were told they would have 20 to 30 minutes to complete the task and then told to begin. During the task, researchers did not interact with the participant unless there was an issue. If the participants finished the task before 25 minutes, they were told to review their work and polish it further until 25 minutes had elapsed for that task. Immediately following a first task, the Flow Short Scale (FSS) was administered on paper so that participants can rate how well they achieved the subjective characteristics of flow states. After completing the FSS, another rest period followed where participants watched another slide show of nature photos silently for 3 minutes. This same procedure was repeated for the second and third tasks, which were followed by the FSS and a rest period. Participants remained seated the entire time and did not eat or drink anything during the data collection.

After the final rest period after the third task, in the post-interview, the researcher reviewed the FSS ratings across the three tasks with the participant and allowed the participant to calibrate and adjust any ratings from earlier tasks after experiencing all three conditions. Care was taken not to unduly influence the participant's ratings, as the researcher mostly summarized the ratings across the conditions in each flow characteristic to verify that the relative differences (or similarities) indeed reflected what the participant felt. The researcher also asked participants to subjectively describe how focused, concentrated, or challenged they felt when performing the tasks. Naturally, the discussions led to comparisons across conditions. The post interview lasted approximately 10-15 minutes, after which the participants were thanked and dismissed.  

\subsection{Physiological Sensors and Measures} \label{Physiological Sensors and Measures}
We selected a set of physiological sensors and measures that we expect to be related to different focus states. The choice of sensors and measures are based on a) findings from prior work in flow from other domains such as gaming and driving (see Section \ref{Background}), and b) constraints of the workplace for sensors in terms of ease of access, comfort level, unobtrusiveness and social acceptance. We ruled out the use of EEG and EKG measuring sensors as well as chest straps as these are either too uncomfortable or not socially acceptable. We also considered webcam-based techniques for measuring a worker's pulse, respiration, affect, or alertness, but having an always-on camera that is often used for meetings and broadcasts can be considered too invasive even for the workplace. 

Previous studies show that heart rate (HR), heart rate variability (HRV), skin conductance, and pupil dilation are related to flow-like states, arousal, and cognitive load (see Sections \ref{Physiological Correlates of Flow} and \ref{Cognitive Load and Stress}). For our study, we settled on using the Kyto ear clip sensor \cite{Bluetoot35:online} which attaches to the user's ear lobe and uses photoplethysmography (PPG) to provide a reliable measure of heart rate and interbeat intervals for quantifying metrics of heart rate variability. The ear clip was not the ideal form factor as it can be conspicuous, but it is close in function to less obtrusive wristbands like the Empatica E4 band \cite{empatica} that can accurately capture interbeat intervals. Participants also wore a Microsoft Band 2 wristband \cite{Microsof12:online} to measure skin conductance as well as heart rate and interbeat intervals. The Kyto ear clip provided an interbeat interval signal with less noise when compared with the MS Band 2 as the latter suffered from motion artifacts while the users were typing and performing the tasks. The MS Band 2 reported almost every heart beat and skin conductance at 5Hz. The Kyto ear clip, it turns out, reported approximately every other interbeat interval, though the signal was much more stable than from the Band 2. Subsequent benchmark tests with an Empatica E4 (not used in this study) showed it had a similar rate of capture of interbeat intervals as the Kyto ear clip.

We also used a Tobii X120 eye tracker \cite{TobiiGam31:online} to track the diameters of the right and left pupils of the participants. The Tobii device was placed below the monitor and angled towards the participant's face. The position was adjusted by the researcher based on the participant's height and comfortable seating position. Calibration of the eye tracker was not necessary because we did not track where the user was looking on the screen, as this is highly task dependent rather then dependent on the participant's focus state. Occasionally, participants might lean in too far towards the monitor to read small text, effectively moving out of the eye tracker's range. In this case, the researcher, who was monitoring the tracking status on a separate display, prompted the participant to lean back slightly and increase the font size on the document if they were having difficulty reading. The Tobii X120 tracked the pupil size at 60Hz.

We wrote custom software to collect the data from Kyto ear clip. Data from the clip was transmitted to a laptop via Bluetooth and a timestamp token was appended to it and saved. For the MS Band 2 data, we used the 'Companion for Band' app from the Android Play Store. We used custom software that relied on the Tobii eye tracker API to collect and log data from the eye tracker to a laptop. During each trial of the study, the researcher manually started the data collection process from these three sensors. The data streams from multiple sensors were time synchronized during the analysis stage by cropping the data before and after the beginning of each task or rest period. Before each trial, special care was taken to synchronize the system clocks of all the sensors and recording devices w.r.t the atomic clock. 
\section{Model Building}
In this section, we describe our approach for preparing the data, training models to predict focus states from physiological parameters, and evaluating the performance of the models. We begin with cleaning the noisy data from the sensors, labeling the data appropriately, extracting features from sensor data, and training the machine learning model to predict different focus states from the training data. 

\subsection{Data Preparation/Cleaning}
First, we cropped data across the multiple sensors to ensure it represented the actual duration of the task performed by the participant. The data clipping range was maximum of start time and the minimum of end time across all sensors for a particular task. This process was repeated for each task and for every participant.

After ensuring the timestamps of the data streams were in sync, we took appropriate measures to clean the noisy data collected from the physiological sensors. We removed ectopic beats from the RR interval data from Kyto ear clip sensor by following the standard clinical practice of removing RR intervals that differ more than 20\% from the preceding interval. For the pupil diameter data, we removed instances where only one eye was visible to the tracker, as well as instances where the right and left pupil diameters differed by more then 10\%. The mean of the left and right pupil diameters was calculated. To prepare the skin resistance data from the MS Band 2 for peak detection, the values were converted to measures of conductance (microSiemens) by taking the reciprocal, then it was linearly upsampled from 5Hz to 8Hz and a low-pass Butterworth filter (order: 6, $W_n$=0.25) was applied to remove values out of bounds. To account for personal differences in physiology, the values for heart rate, pupil diameter, and skin conductance data were standardized within each participant by converting each value to a z-score based on the means for each participant.

\subsection{Feature Extraction}
After data cropping and cleaning, higher level features were extracted from the data to create training and testing samples for machine learning. We divided the time series data from the sensors into windows, with each window corresponding to a training/test sample and labeled with their corresponding condition for machine learning. The windows allowed us to calculate aggregated measures for that time windows. To calculate metrics for HRV, we need multiple RR intervals to measure their variability. To calculate frequency-domain features of HRV, we need at least 64 RR intervals, which formed the lower bound for the window size. The upper bound was found to be 120, when any higher would result in insufficient samples for training and evaluating focus state classifiers. Further, windows could overlap with each other to create more samples. Treating window size and overlap as a hyperparameters in a grid search, the optimal window size was found to be 90 with an overlap of 30. Thus we selected window sizes to be determined by the time spanned by 90 consecutive RR intervals. Data from the other sensor streams (pupil and skin conductance) between the timestamp of the first and last RR interval in the window were retrieved and associated with this window. Based on the data for each window, we calculate features to characterize the heart rate, heart rate variability, skin conductance, and pupil diameter. To scale values so that they would be comparable across multiple participants, the newly derived features were then standardized within each participant by converting it to a z-score based on the mean value for that participant. Overall, our feature set (Table \ref{table:1}) consists of 770 windows of data, each with a total of 29 features, 23 for HRV, 3 for skin conductance, and 2 for pupil size.

\begin{table}[h!]
\begin{center}
\begin{tabular}{ |l|l|p{9cm}| } 
\hline
Category & Feature & Definition \\
\hline
\multirow{15}{6em}{HRV (Time domain)}
& maxHR &  maximum heart rate (bpm)\\ 
& minHR &  maximum heart rate (bpm)\\
& mHR &  average heart rate (bpm)\\
& sdHR &  standard deviation of HR (bpm)\\
& mRRi &  average RR interval length (ms) \\
& nn5 &  number of successive RR intervals that differ more than 5ms \\
& nn10 &  number of successive RR intervals that differ more than 10ms \\
& nn20 &  number of successive RR intervals that differ more than 20ms \\
& nn50 &  number of successive RR intervals that differ more than 50ms \\
& pnn5 &  percentage of successive RR intervals that differ more than 5ms \\
& pnn10 &  percentage of successive RR intervals that differ more than 10ms \\
& pnn20 &  percentage of successive RR intervals that differ more than 20ms \\
& pnn50 &  percentage of successive RR intervals that differ more than 50ms \\
& sdNN &  standard deviation of RR interval lengths \\
& rmssd &  root mean square of successive RR interval differences \\
\hline
\multirow{6}{6em}{HRV (Frequency domain)} 
& LF & low frequency (RR fluctuations within 0.04-0.15 Hz) \\ 
& HF & high frequency (RR fluctuations within 0.15-0.4 Hz) \\ 
& vLF & very low frequency (RR fluctuations within 0-0.04 Hz) \\ 
& LF/HF & ratio of LF and HF \\ 
& LFnu & LF/(LF+HF) \\ 
& total\_power & total power across all frequencies \\ 
\hline
\multirow{2}{6em}{HRV (Non-linear)}
& SD2 & length of Poincaré plot, a measure of continuous long-term variability  \\
& SD1/SD2 & ratio of short-term and long-term variability of HRV \\
\hline
\multirow{3}{6em}{Skin Conductance} 
& EDA & mean of skin conductance in microSiemens \\ 
& EDA\_variance & variance of skin conductance in microSiemens \\ 
& peaks & number of detected peaks, or electro-dermal responses, as defined in \cite{taylor2015automatic} \\ 
\hline
\multirow{2}{6em}{Pupil} 
& pupil\_diameter\_mean & mean of the pupil diameters  \\ 
& pupil\_diameter\_variance & variance of the pupil diameters \\ 
\hline
\end{tabular}
\end{center}
\caption{Features extracted from each window of data across the sensor streams}
\label{table:1}
\end{table}

\subsection{Training and Evaluating Models} \label{Training and Evaluating Models}
We used the Python Scikit-learn framework for training and evaluating machine learning models. We applied different learning approaches including SVM (with a linear kernel), Decision Tree (using Gini as the split criterion), and Random Forest (number of estimators=10). We trained models for the following classification tasks: 
\begin{enumerate}
\item Non-flow vs Automatic Flow vs Balanced flow //classifying all three focus states
\item Non-flow vs Flow states //identifying a non-flow state from either kinds of flow state
\item Balanced Flow vs Other Work States //identifying the ideal balanced flow from other working states
\item Rest states vs Working States //distinguishing between not working vs working
\end{enumerate}

The pipeline begins with feature selection to select the top 20\% of features that ranked highly according to a univariate ANOVA test. To approximate more generalized performance, models were trained and evaluated using leave-one-participant-out cross validation where one participant's data was completely left out of the training set and used only for evaluation in the test set. Scores were averaged across all the participants for an overall performance score. Care was taken to not include of the participant's own data in the training set, to allow the results to be as generalized (not person-specific) as possible. For classification cases in which the labeled samples are imbalanced, we used the imblearn python package to balance the dataset by undersampling the majority class, randomly selecting a subset of majority class samples to equal the number of samples in the minority class. With this balanced set, a model was trained and evaluated. This procedure repeated 10 times with a new random subset of the majority class and the average accuracy scores computed. We also tried SMOTE oversampling and had similar results to undersampling, so we report the results of the more conservative undersampling approach that does not require the extra step of generating synthetic data. For binary classification, we report AUC (area under the receiver operating characteristic curve) for the performance metric because it provides a captures the tradeoffs of true and false positives. For multi-class classification, we report the  weighted-F1 metric.



\section{Results}
Based on the data collected from the participants performing three different work tasks that corresponded to three different focus states induced in the experimental manipulation, we report on 1) a manipulation check of how well each task actually induced the intended focus state in the participant 
and 2) the performance of a machine learning models that we built (see \ref{Training and Evaluating Models}) using combinations of features from the physiological data to predict the focus state of the participant. 

\subsection{Manipulation Check} \label{Manipulation Check}
\begin{figure}
\centering
  \includegraphics[width=1\columnwidth]{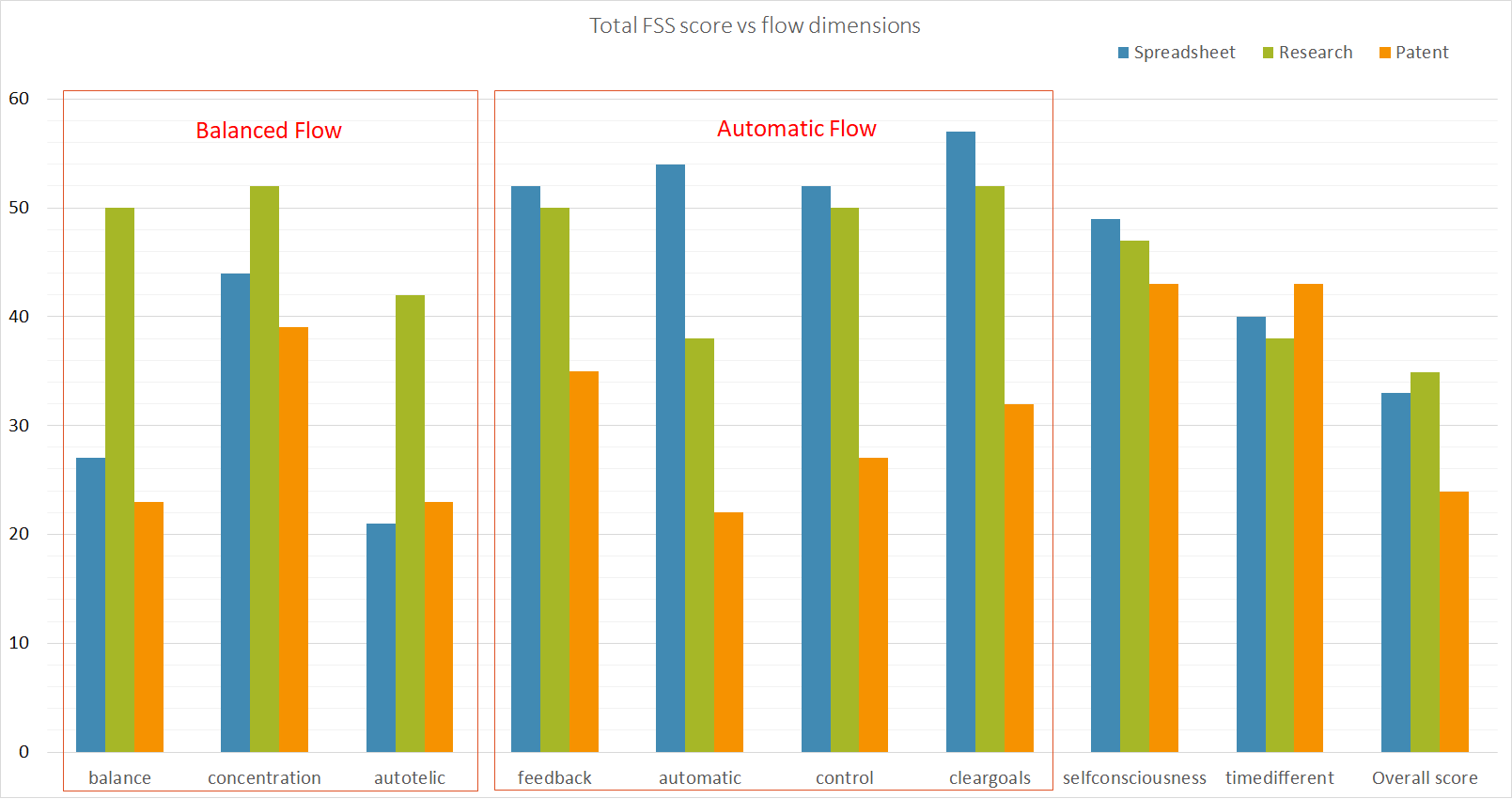}
  \caption{Ratings of the characteristics of flow from the Flow Short Scale. The composite overall score (right) shows that the Patent task was rated as much less flow than the Spreadsheet and Research tasks, which were rated nearly equivalently in the overall score. However, the experience in the Research task rated more highly in the characteristics of Challenge/Skill balance, High concentration, and Autotelic experience, which we characterize as "Balanced Flow". In contrast, the experience in the Spreadsheet task was rated more highly in Unambiguous feedback, Automaticity, Sense of control, and Clear goals, which we characterize as "Automatic Flow".}
    \label{fig:manipulation-chart}
\end{figure}

Flow states are highly dependent on the interaction effect between the skills of the individual and work they are performing. It is impossible to design a set of realistic tasks that would induce flow states in all people. Therefore, we designed three tasks that would induce different focus states in specific type of knowledge worker in our study, researchers. To verify the level of flow experienced in each task, participants completed the Flow Short Scale (FSS) \cite{vollmeyer_motivational_2006} immediately after each task. For a composite measure of flow, we summed up the responses across the 9 items in the FSS, for a maximum score of 45. The higher the score, the greater sense of flow reported by the participant. The results of the manipulation check is shown in Figure \ref{fig:manipulation-chart}.

The Research task was designed to induce the highest sense of flow, compared with the Spreadsheet task and Patent task. We would thus expect for the composite FSS flow score to be the highest in the Research task than the other tasks. However, on average across participants, the composite FSS flow score for the Research task (mean=34.9, sd=2.9) was only statistically significantly higher than the flow score in the Patent task (mean=23.9, sd=5.1) (F(2,22)=34.5, p<0.0001), but not significantly higher than in the Spreadsheet task (mean=33.0, sd=4.6), as shown in Figure \ref{fig:manipulation-chart}, rightmost columns. In other words, participants did not experience as much focused flow in the Patent task as we expected, but rather, the overall amount of focused flow in the Research and Spreadsheet tasks were roughly equivalent, which is different than what we intended with our experimental manipulation. However, looking more closely at the individual items in the FSS, we noticed that the Research and Spreadsheet tasks were rated differently. For the Research task, the FSS items that were rated significantly higher than other tasks include: challenge/skill balance, effortless concentration, and autotelic experience, F[1,11]=24.6, p=0.0004. Participants were able to get into a sense of flow in the Research task because it was interesting and enjoyable to read/summarize a research paper of their own choosing which provided a meaningful level of challenge. For the Spreadsheet task, the FSS items that were rated highly (in fact, higher than in the Research task) include: feedback, automaticity, sense of control, and clear goals, F[1,11]=8.6, p=0.014. Surprisingly, in the Spreadsheet task, participants got a sense of flow when performing the too-easy, almost mindless task of spell-checking email addresses because the goals were clear, they felt in control, and they could perform the task almost automatically. Even though we purposely tried to reduce the sense of control by restricting the use of macros, search-and-replace, and other faster ways to complete the task, participants still experienced a high level of control just using the keyboard and mouse to manually edit each cell. Even though it was unintended, it is not surprising, in retrospect, that editing each cell in a single-column spreadsheet can be done repetitively at a rhythm with automatic actions. 

We also verified that the difficulty of the tasks were as we expected using the responses to an additional question on the FSS asking participants to rate the difficulty of the task relative to their regular tasks from 1 (easy) to 9 (difficult). The Patent task (mean=8.6, sd=0.7) was rated significantly more difficult than the Research task (mean=4.7, sd=1.9) which was rated significantly more difficult than the Spreadsheet task (mean=1.75, sd=0.9). 

To summarize the effectiveness of the experimental manipulation, we interpret the Patent task as inducing a non-flow state, as we had expected in which participants rated the 9 characteristics of flow in the FSS to be rather low, consistent with the qualitative feedback from the participants. The other two tasks, Research and Spreadsheet, can be thought of two types of flow-like states, which we call "balanced flow" and "automatic flow," respectively. The Research task induced a sense of "balanced flow," by being both enjoyable and meaningful, and balancing skill with challenge. The Spreadsheet task inducing a sense of "automatic flow" by facilitating automatic actions, being clearly defined, and giving a good sense of control, despite having a poor challenge/skill balance. Instead of three conditions corresponding to boredom, flow, and anxiety, our manipulation resulted in verified conditions of two types of flow and one non-flow: automatic flow, balanced flow, and anxiety (non-flow). In the next section, we investigate how these three states differed in terms of the participant's physiological signals.


\subsection{Machine learning model performance}
We trained and evaluated models to classify the different focus states, including three-way and two-way classifications that can be used in different applications. Among the algorithms we tried, SVM (with a linear kernel) almost universally produced the best performance, so we focus on reporting results from SVM models. See Table \ref{table:2}.

\begin{table}[h]
\begin{center}
\begin{tabular}{ |l||c|c|c|  }
 \hline
 Classifier & AUC (avg) & (sd)\\
 \hline
 Non-flow vs Flow state & 0.889 & 0.087\\
 Balanced Flow vs Other Work & 0.772 & 0.176\\
 Rest vs Work    & 0.980 & 0.040\\
 Non-flow vs Automatic Flow vs Balanced Flow & 0.668 (F1) & 0.200 \\
 \hline
\end{tabular}
\end{center}
\caption{Performance of classifiers for each decision task. Note: The AUC (Area Under the Receiver Operator Curve) is used for binary classifiers. The weighted-F1 metric is used for multi-class classification. All metrics were averaged across all participants in the leave-one-subject-out cross validation. The standard deviation across participants is also reported.}
\label{table:2}
\end{table}

\subsubsection{Non-flow vs Automatic Flow vs Balanced flow}
We first report the results of a three-way classification of the three different focus states, non-flow (from our patent task), automatic flow (from the spreadsheet task), and balanced flow (from the research task). The SVM-based model gave a weighted-F1 score averaged across all participants of 0.668 (sd=0.2). The F1 scores for each of the classes were: 0.70 for automatic flow, 0.60 for balanced flow and 0.70 for non-flow. The top ranked features by univariate ANOVAs include pupil\_diameter\_mean, peaks, sdNN, SD2, maxHR, rmssd, and SD1. 
The three-way classification is important for applications that need to distinguish between all three tasks, non-flow, automatic flow, and balanced flow. Next we consider binary classifications that also have potential applications in the workplace. 

\subsubsection{Non-flow vs Flow states}
In many applications, it is more important to detect when a worker is clearly \emph{not} in a flow-like state, such as when the worker is overworked, anxious, or when the challenge of the task far exceeds the worker's ability. In this case, a binary classifier that can distinguish between a non-flow state from the either of the two flow states (that is, treating both automatic flow and balanced flow both as the same higher-level label of 'flow') would be useful. The SVM-based model for non-flow vs flow states had an average AUC across participants of 0.889 (sd=0.087). The highest ranked features for this case included pupil\_diameter\_mean, sdNN, SD2, rmssd, SD1, and total\_power. 

\subsubsection{Balanced flow vs Other Work}
Balanced flow is particularly useful for understanding when a worker is performing at a productive level, when they are exercising a good amount of their skills to complete a sufficiently challenging task. The SVM-based binary classifier for detecting balanced flow vs other work (automatic flow and not flow) had an AUC of 0.772 averaged across participants (sd=0.176). The highest ranked features for this case included: peaks, total\_power, LF, SD1, and rmssd.


\subsubsection{Rest vs Non-Rest States}
In our experiment, we had participants sit idly and watch a relaxing slide show of nature photos for 3 minutes in between the work tasks. This provides us an opportunity to build a classifier that can distinguish between these rest periods and work sessions. We built a binary classifier with data from the rest periods labeled as rest and the data collected during the tasks. We excluded the initial rest period before the experiment started because it could have affected by external factors such as the physical activity required to travel to our office. After balancing the dataset due to many more samples of work data collected in our study, we get an AUC of 0.98 averaged across participants of (sd=0.04). The highest ranked features being pupil\_diameter\_mean, pupil\_diameter\_variance, pnn20, nn20, nn10, pnn10. 


\section{Discussion}
Detecting a worker's focus state at work can enable individual workers to identify the context of flow and non-flow experiences and organizations to understand the engagement and well-being of their workers. In our study, we induced flow and non-flow states in workers by having them work on tasks with different challenge/skill balances and measured their physiological with workplace-appropriate sensors. 

\subsection{The Construct of Automatic Flow}
A manipulation check of our tasks using the Flow Short Scale revealed that instead of one flow state and two non-flow states, boredom and anxiety, we actually induced one non-flow state (anxiety) and two flow states, balanced flow and automatic flow, that rated highly in different characteristics of flow. Even though these two flow-like states were not what we initially intended, the manipulation revealed that flow at work can take on different forms, especially for adaptable workers who complete a breadth of types of tasks. Participants reported that some of their own work tasks were as routine and seemingly mindless as the spreadsheet task in the experiment that registered as automatic flow and can be a good break from other more difficult work. Some participants reported that during the spreadsheet task, they were engaged in optimizing their keystrokes and movements to complete the task as quickly as possible. They tried different combinations of keystrokes and mousing to find which was the fastest, and only after they reached a stable routine that they could not further improve, did they experience the boredom that we had intended. This optimization behavior might be responsible for unexpected germane cognitive load, that is, mental effort devoted to processing, constructing an efficient schema of task actions, which in fact has a positive correlation with flow experiences \cite{chang2017game}. Thus, participants looked for a challenge in this dull, boring task to stimulate themselves, and when combined with the rhythm actions and clear goals and feedback, this supported a good sense of the automatic type of flow.

To answer our first research question about how to induce flow states by manipulating work task characteristics, our manipulation check shows that we were successful at creating tasks that induced different focus states in our participants (researchers). In particular, the patent reading task was designed to be non-flow, and it proved to be the case because it was too difficult (relative to the skills of our participants), had unclear goals, did not feel automatic, had a low level of control, and was not enjoyable. The task of summarizing a research paper was designed to induce more flow by having a good challenge/skill balance, be enjoyable, and highly motivating to facilitate high concentration and our manipulation check proved that it was. However, designing a truly boring (non-flow) task was more difficult. Tasks with low challenge tend to be simpler tasks that have relatively clear goals, a limited set of actions leading to a fairly high level of control and automatic actions that can contribute to feelings of flow. Furthermore, people are adaptable, and when forced to do a boring, uninteresting task, they find ways to make it challenging or  interesting, as we found when our participants spent effort not only to perform the task but to optimize their actions to be more efficient. One lesson learned is that simply manipulating the challenge/skill balance dimension of a task, which is commonly done in other domains like in games and driving, does not necessarily result in a greater or lower sense of flow. Even unchallenging tasks when performed with clear goals, high sense of control, and with automatic, repetitive actions at a rhythm can still result in a sense of flow-like focus in workers. Overall, we demonstrated it is feasible to use the dimensions of flow from the Flow Short Scale as a guide to tune the parameters of a task to induce flow or non-flow states, but care should be taken when designing low challenge/high skill tasks to limit the amount of automaticity allowed, perhaps with interruptions or changing goals throughout the task. 


\subsection{Detecting Flow States using Physiological Sensing}
Addressing our second research question, we built a machine learning model is able to distinguish the too-challenging non-flow state from the flow states (automatic and balanced), at roughly 0.889 AUC on average across participants. We used a leave-one-participant-out cross validation approach, so this best approximates the performance of a generalized model that does not include any personalization to the individual. It is possible that including data from a calibration session for each user would improve the performance of the classifier by modeling some of the physiological responses unique to each individual. Our model performance for classifying flow states is slightly better to similar physiology-based approaches such as Müller et al.'s \cite{muller2015stuck} 71\% accuracy in classifying worker's emotional valence and to Fritz et al.'s \cite{fritz_using_2014} 65\% accuracy in classifying task difficulty. Ara et al. \cite{ara2009predicting} also found some significant correlations (|r|>0.2) between motion frequency and reported flow states, but however, their results show significant variations across participants in which specific motion frequency range is related to flow as well as the sign of the correlation. Thus, their approach would need to be personalized to each worker's frequency range for an effective flow state classifier. 






\subsection{Study Limitations}
Our study had a few limitations and trade-offs. One of the most influential decisions we made was to collect data in a controlled lab setting rather than gather data from worker's actual workday in the field. In the lab, we had to construct tasks that would induce different flow states, but may not have represented the breadth of tasks that workers perform in their daily work. However, more important than the type of task themselves are the perceived flow states that the tasks induced and experienced by the participant. We were able to verify that we had three distinct kinds of focus states by using the Flow State Scale as a manipulation check. All the features used in our machine learning models were not specific to any task but rather a physiological response to the task demands. Furthermore, a related limitation is that throughout each 25-minute task, the participant may drift in and out of varying degrees of focus, yet we associate all the data from that session with the same focus state label. When designing the study, we considered having participants periodically self-report their flow state during the task, but we decided that any sort of prompt or extraneous action may disrupt their ability to enter or maintain their focus state. After each task, we asked participants if they noticed themselves feeling more or less focused in the beginning, middle, or end of the task, or if they experienced any mind-wandering. None reported any significant changes in feelings as the task proceeded, and the manipulation check showed that as a whole, each task induced a different focus state from the others. 
We are eager to take our future research into the field with more participants from different occupations and expertise levels and collect data from workers throughout their work day when they experience the levels of flow not easily induced in the lab. 

Another limitation of our study is in the specific set of sensors we used. Even though we used higher-level signals such as HRV, skin conductance, and pupil size derived from the raw sensor data, our models likely took into account the biases and noise introduced from the specific sensors we used, and thus using another kind of heart rate sensor or pupil size sensor (such as the simple, webcam-based pyGaze \cite{pygaze}) might yield different results. 

\subsection{Potential Applications}
Having the ability to detect when a worker is in a flow-like state or not can be useful in a number of applications. In this section, we sketch some of these applications to inspire future opportunities for research. Broadly, these applications fall into two classes: for individuals and for organizations. 

Flow sensing for the individual can enable individual workers to take a personal informatics approach to identify when they are in flow or not and improve their productivity. Combined with other data streams such as time, meeting schedules, sleep, physical activity, or tasks completed, an individual work can identify the contextual factors that contribute to flow or non-flow states. For example, a worker can identify that she is more likely to enter in a flow state on days after a good night sleep (due to restfulness) or shortly before meetings (due to the clarity of goals). She can adjust her schedule to prioritize downtime or schedule dedicated work time in between meetings to boost opportunities for more flow experiences which can improve her productivity and enjoyment of her work. Making the system a little more proactive, the system can not only collect data and reveal trends, but also can be a flow coach, providing appropriate interventions to facilitate more flow experiences. For example, when detecting that the worker is in non-flow and is feeling overworked or anxious for a while, the flow coach can prompt the worker to switch temporarily to a less challenging task to reduce her stress level or suggest some brief physical activity \cite{Choi:2016:EUE:2971648.2971756} or a dose of blue light found to be effective for increasing creativity by Abdullah et al. \cite{Abdullah:2016:SLC:2971648.2971751}. The flow coach, given its access to the worker's biometrics, can be combined with biofeedback techniques described by Ghandeharioun \& Picard \cite{Ghandeharioun:2017:BEI:3027063.3053164} and Costa et al. \cite{Costa:2016:ELB:2971648.2971752} to automatically regulate emotions and reduce anxiety using (false) feedback. 

Interruption management systems can also be augmented to take into account the worker's flow state. When in a flow state, people generally are good at screening out temporary distractions and concentrating at the task at hand. Thus, interruption managers can deliver short notifications, even those that might require the worker to perform an quick action, when the worker is strongly in flow because resuming the main task will likely be easier than when not in flow. Similarly, website blockers can release their lock when the system detects that the worker has reached an adequate level of flow and is less susceptible to self-distractions, allowing the worker to use previously-blocked websites to get work done. 

Detecting when workers are in flow or not flow can also be useful to employers. Systems can monitor a worker's flow states and their associated tasks. One of the main factors that facilitates flow experiences is a good match between the worker's skills and the challenge of the task. Thus tasks that are consistently associated with low or no flow in worker may either be too difficult or too easy for the worker. Organizations can use continuous flow sensing to identify which workers are bored and assign them more challenging responsibilities and also identify which workers are anxious and provide additional training to support improve their skills. By better matching the employee's skills to the right tasks or job function can not only improve output but also support employees' wellbeing and engagement. Of course, care should be taken when considering how the internal focus states of employees might be used in performance evaluations so as not to unduly mischaracterize their abilities or contributions.

Organizations can also use data about the focus state of employees as a proxy for production metrics such as likelihood of making a mistake. For example, factory workers who work on an assembly line might make more mistakes when they are in strongly non-flow states because the worker is either inadequately skilled, may be distracted due to boredom, or for a temporary reason. Thus, the flow states of the workers can give an indication of the production quality of a component, and additional quality assurance can be assigned to production lots when workers were feeling less focused. Combining flow state data with activity recognition of factory workers \cite{Maekawa:2016:TPF:2971648.2971721} can provide further insight how well workers have mastered their task.

\subsection{Future Work}
We plan to conduct a field study to evaluate our flow state classifier technique with data outside the lab and with workers of varying backgrounds and expertise perform a wider variety of tasks. In addition, we are interested in developing capabilities to quantify not only the type of flow but also the degree of flow a worker is experiencing. Having this additional information can help systems know better how to interact with workers to guide them to enter into and maintain flow experiences at work. 
\section{Conclusion}
Flow experiences help workers be more productive, enjoy their work, and feel more engaged at work. In this paper, we investigated how workers' physiological signals changed in flow and non-flow states while performing work tasks in a lab study. We used the nine characteristics of flow experiences to design work tasks that induced different types of focus states. A manipulation check revealed that participants experienced one non-flow state (in a task too difficult relative the worker's skills), and two types of flow states, balanced flow (in a task that had a good challenge/skill balance and enjoyment) and automatic flow (in an easy task with clear goals and automatic actions). A statistical analysis showed that compared with the other focus states, the balanced flow state is most distinguished by the highest number of skin conductance peaks whereas the automatic flow state is characterized by the lowest variation in measures of skin conductance. The overly challenging non-flow state was associated with highest heart rate variability (rmssd, sdNN, total\_power), suggesting that our participants were able to recruit some parasympathetic activity to calm themselves and help them focus on the difficult task. We trained and evaluated non-personalized machine learning classifiers to classify the three focus states with 67\% average accuracy across participants, not-flow vs flow states with 75\% accuracy, balanced flow vs automatic flow with 74\% accuracy, and rest vs non-rest with 94\% accuracy. Overall, our approach in using physiological signals to characterize and predict flow states compares favorably to existing approaches for detecting flow and emotional valence in workers. Our results are a promising initial step in helping workers and organizations better track flow and non-flow experiences at work to support better productivity, task fit, and well-being for workers.

\section*{Acknowledgements}
We thank Sumeet Jain for playing an integral role in building the infrastructure and helping to run the data collection study. We also thank Mitesh Patel for helpful advice on machine learning. We thank Yanxia Zhang and Daniel Avrahami for their useful discussions.

\bibliographystyle{unsrt}
\bibliography{references}

\begin{thebibliography}{10}

\bibitem{demerouti2006job}
Evangelia Demerouti.
\newblock Job characteristics, flow, and performance: The moderating role of
  conscientiousness.
\newblock {\em Journal of occupational health psychology}, 11(3):266, 2006.

\bibitem{fredrickson2004broaden}
Barbara~L Fredrickson.
\newblock The broaden-and-build theory of positive emotions.
\newblock {\em Philosophical Transactions of the Royal Society B: Biological
  Sciences}, 359(1449):1367, 2004.

\bibitem{lefevre1988flow}
Judith LeFevre.
\newblock Flow and the quality of experience during work and leisure.
\newblock In {\em Optimal experience: {Psychological} studies of flow in
  consciousness}, pages 307--318. Cambridge University Press, New York, NY, US,
  1988.

\bibitem{stress_workplace_2011}
The American~Institute of~Stress.
\newblock Workplace stress.
\newblock \url{https://www.stress.org/workplace-stress/}.

\bibitem{kass2001watching}
Steven~J Kass, Stephen~J Vodanovich, Claudia~J Stanny, and Tiffany~M Taylor.
\newblock Watching the clock: Boredom and vigilance performance.
\newblock {\em Perceptual and motor skills}, 92(3\_suppl):969--976, 2001.

\bibitem{van2014boredom}
Madelon~LM van Hooff and Edwin~AJ van Hooft.
\newblock Boredom at work: Proximal and distal consequences of affective
  work-related boredom.
\newblock {\em Journal of occupational health psychology}, 19(3):348, 2014.

\bibitem{cranston2013}
Susie Cranston and Scott Keller.
\newblock Increasing the ‘meaning quotient’ of work {\textbar} {McKinsey}
  \& company.
\newblock
  \url{https://www.mckinsey.com/business-functions/organization/our-insights/increasing-the-meaning-quotient-of-work},
  2013.

\bibitem{peifer2014relation}
Corinna Peifer, Andr{\'e} Schulz, Hartmut Sch{\"a}chinger, Nicola Baumann, and
  Conny~H Antoni.
\newblock The relation of flow-experience and physiological arousal under
  stress—can u shape it?
\newblock {\em Journal of Experimental Social Psychology}, 53:62--69, 2014.

\bibitem{tozman2015understanding}
Tahmine Tozman, Elisabeth~S Magdas, Hamish~G MacDougall, and Regina Vollmeyer.
\newblock Understanding the psychophysiology of flow: A driving simulator
  experiment to investigate the relationship between flow and heart rate
  variability.
\newblock {\em Computers in Human Behavior}, 52:408--418, 2015.

\bibitem{nakamura2014concept}
Jeanne Nakamura and Mihaly Csikszentmihalyi.
\newblock The concept of flow.
\newblock In {\em Flow and the foundations of positive psychology}, pages
  239--263. Springer, 2014.

\bibitem{massimini_systematic_1988}
Fausto Massimini and Massimo Carli.
\newblock The systematic assessment of flow in daily experience.
\newblock In {\em Optimal experience: Psychological studies of flow in
  consciousness}, pages 266--287. Cambridge University Press, 1988.

\bibitem{csikszentmihalyi1989optimal}
Mihaly Csikszentmihalyi and Judith LeFevre.
\newblock Optimal experience in work and leisure.
\newblock {\em Journal of personality and social psychology}, 56(5):815, 1989.

\bibitem{bakker2005flow}
Arnold~B Bakker.
\newblock Flow among music teachers and their students: The crossover of peak
  experiences.
\newblock {\em Journal of vocational behavior}, 66(1):26--44, 2005.

\bibitem{fullagar2009flow}
Clive~J Fullagar and E~Kevin Kelloway.
\newblock Flow at work: An experience sampling approach.
\newblock {\em Journal of occupational and organizational psychology},
  82(3):595--615, 2009.

\bibitem{ilies2017flow}
Remus Ilies, David Wagner, Kelly Wilson, Lucia Ceja, Michael Johnson, Scott
  DeRue, and Dan Ilgen.
\newblock Flow at work and basic psychological needs: Effects on well-being.
\newblock {\em Applied Psychology}, 66(1):3--24, 2017.

\bibitem{engeser2016fluctuation}
Stefan Engeser and Nicola Baumann.
\newblock Fluctuation of flow and affect in everyday life: A second look at the
  paradox of work.
\newblock {\em Journal of Happiness Studies}, 17(1):105--124, 2016.

\bibitem{gaggioli2013psychophysiological}
Andrea Gaggioli, Pietro Cipresso, Silvia Serino, and Giuseppe Riva.
\newblock Psychophysiological correlates of flow during daily activities.
\newblock {\em Annual Review of Cybertherapy and Telemedicine}, 191:65--69,
  2013.

\bibitem{de2010psychophysiology}
{\"O}rjan De~Manzano, T{\"o}res Theorell, L{\'a}szl{\'o} Harmat, and Fredrik
  Ull{\'e}n.
\newblock The psychophysiology of flow during piano playing.
\newblock {\em Emotion}, 10(3):301, 2010.

\bibitem{ullen201010}
Fredrik Ull{\'e}n, {\"O}rjan de~Manzano, T{\"o}res Theorell, and L{\'a}szl{\'o}
  Harmat.
\newblock 10 the physiology of effortless attention: Correlates of state flow
  and flow proneness.
\newblock {\em Effortless attention: A new perspective in the cognitive science
  of attention and action}, page 205, 2010.

\bibitem{jaeggi2008improving}
Susanne~M Jaeggi, Martin Buschkuehl, John Jonides, and Walter~J Perrig.
\newblock Improving fluid intelligence with training on working memory.
\newblock {\em Proceedings of the National Academy of Sciences},
  105(19):6829--6833, 2008.

\bibitem{rowe1998heart}
Dennis~W Rowe, John Sibert, and Don Irwin.
\newblock Heart rate variability: Indicator of user state as an aid to
  human-computer interaction.
\newblock In {\em Proceedings of the SIGCHI conference on Human factors in
  computing systems}, pages 480--487. ACM Press/Addison-Wesley Publishing Co.,
  1998.

\bibitem{cinaz2013monitoring}
Burcu Cinaz, Bert Arnrich, Roberto La~Marca, and Gerhard Tr{\"o}ster.
\newblock Monitoring of mental workload levels during an everyday life
  office-work scenario.
\newblock {\em Personal and ubiquitous computing}, 17(2):229--239, 2013.

\bibitem{dong2011driver}
Yanchao Dong, Zhencheng Hu, Keiichi Uchimura, and Nobuki Murayama.
\newblock Driver inattention monitoring system for intelligent vehicles: A
  review.
\newblock {\em IEEE transactions on intelligent transportation systems},
  12(2):596--614, 2011.

\bibitem{bailey2008understanding}
Brian~P Bailey and Shamsi~T Iqbal.
\newblock Understanding changes in mental workload during execution of
  goal-directed tasks and its application for interruption management.
\newblock {\em ACM Transactions on Computer-Human Interaction (TOCHI)},
  14(4):21, 2008.

\bibitem{klingner2010measuring}
Jeffrey~Michael Klingner.
\newblock {\em Measuring cognitive load during visual tasks by combining
  pupillometry and eye tracking}.
\newblock PhD thesis, Stanford University, 2010.

\bibitem{wang2013indexing}
Weihong Wang, Zhidong Li, Yang Wang, and Fang Chen.
\newblock Indexing cognitive workload based on pupillary response under
  luminance and emotional changes.
\newblock In {\em Proceedings of the 2013 international conference on
  Intelligent user interfaces}, pages 247--256. ACM, 2013.

\bibitem{chang2017game}
Chi-Cheng Chang, Chaoyun Liang, Pao-Nan Chou, and Guan-You Lin.
\newblock Is game-based learning better in flow experience and various types of
  cognitive load than non-game-based learning? perspective from multimedia and
  media richness.
\newblock {\em Computers in Human Behavior}, 71:218--227, 2017.

\bibitem{ara2009predicting}
Koji Ara, Nobuo Sato, Satomi Tsuji, Yoshihiro Wakisaka, Norio Ohkubo, Youichi
  Horry, Norihiko Moriwaki, Kazuo Yano, and Miki Hayakawa.
\newblock Predicting flow state in daily work through continuous sensing of
  motion rhythm.
\newblock In {\em Networked Sensing Systems (INSS), 2009 Sixth International
  Conference on}, pages 1--6. IEEE, 2009.

\bibitem{bucher2017flow}
Eliane Bucher and Christian Fieseler.
\newblock The flow of digital labor.
\newblock {\em new media \& society}, 19(11):1868--1886, 2017.

\bibitem{mauri2011facebook}
Maurizio Mauri, Pietro Cipresso, Anna Balgera, Marco Villamira, and Giuseppe
  Riva.
\newblock Why is facebook so successful? psychophysiological measures describe
  a core flow state while using facebook.
\newblock {\em Cyberpsychology, Behavior, and Social Networking},
  14(12):723--731, 2011.

\bibitem{muller2015stuck}
Sebastian~C M{\"u}ller and Thomas Fritz.
\newblock Stuck and frustrated or in flow and happy: Sensing developers'
  emotions and progress.
\newblock In {\em Software Engineering (ICSE), 2015 IEEE/ACM 37th IEEE
  International Conference on}, volume~1, pages 688--699. IEEE, 2015.

\bibitem{mark2014bored}
Gloria Mark, Shamsi~T Iqbal, Mary Czerwinski, and Paul Johns.
\newblock Bored mondays and focused afternoons: the rhythm of attention and
  online activity in the workplace.
\newblock In {\em Proceedings of the SIGCHI Conference on Human Factors in
  Computing Systems}, pages 3025--3034. ACM, 2014.

\bibitem{vollmeyer_motivational_2006}
Regina Vollmeyer and Falko Rheinberg.
\newblock Motivational {Effects} on {Self}-{Regulated} {Learning} with
  {Different} {Tasks}.
\newblock {\em Educational Psychology Review}, 18(3):239--253, September 2006.

\bibitem{Bluetoot35:online}
Kyto.
\newblock Bluetooth mobile hrv heart rate monitor with ear clip and fingertip
  sensor – kyto fitness technology.
\newblock
  \url{https://kytofitness.com/collections/heart-rate-monitor/products/bluetooth-mobile-heart-rate-monitor-with-ear-clip-kyto2935}.

\bibitem{empatica}
Empatica.
\newblock Real-time physiological signals e4 {EDA}/{GSR} sensor.

\bibitem{Microsof12:online}
Microsoft Band.
\newblock Microsoft band | official site.
\newblock \url{https://www.microsoft.com/en-us/band}.

\bibitem{TobiiGam31:online}
Tobii.
\newblock Tobii gaming | eye trackers for pc games in desktop, laptops \&
  monitors.
\newblock \url{https://tobiigaming.com/products/}.

\bibitem{taylor2015automatic}
Sara Taylor, Natasha Jaques, Weixuan Chen, Szymon Fedor, Akane Sano, and
  Rosalind Picard.
\newblock Automatic identification of artifacts in electrodermal activity data.
\newblock In {\em Engineering in Medicine and Biology Society (EMBC), 2015 37th
  Annual International Conference of the IEEE}, pages 1934--1937. IEEE, 2015.

\bibitem{fritz_using_2014}
Thomas Fritz, Andrew Begel, Sebastian~C. Müller, Serap Yigit-Elliott, and
  Manuela Züger.
\newblock Using psycho-physiological measures to assess task difficulty in
  software development.
\newblock In {\em Proceedings of the 36th International Conference on Software
  Engineering}, {ICSE} 2014, pages 402--413. {ACM}, 2014.

\bibitem{pygaze}
Pygaze.
\newblock Open source eye-tracking software and more.
\newblock \url{http://www.pygaze.org/}.

\bibitem{Choi:2016:EUE:2971648.2971756}
Woohyeok Choi, Aejin Song, Darren Edge, Masaaki Fukumoto, and Uichin Lee.
\newblock Exploring user experiences of active workstations: A case study of
  under desk elliptical trainers.
\newblock In {\em Proceedings of the 2016 ACM International Joint Conference on
  Pervasive and Ubiquitous Computing}, UbiComp '16, pages 805--816, New York,
  NY, USA, 2016. ACM.

\bibitem{Abdullah:2016:SLC:2971648.2971751}
Saeed Abdullah, Mary Czerwinski, Gloria Mark, and Paul Johns.
\newblock Shining (blue) light on creative ability.
\newblock In {\em Proceedings of the 2016 ACM International Joint Conference on
  Pervasive and Ubiquitous Computing}, UbiComp '16, pages 793--804, New York,
  NY, USA, 2016. ACM.

\bibitem{Ghandeharioun:2017:BEI:3027063.3053164}
Asma Ghandeharioun and Rosalind Picard.
\newblock Brightbeat: Effortlessly influencing breathing for cultivating
  calmness and focus.
\newblock In {\em Proceedings of the 2017 CHI Conference Extended Abstracts on
  Human Factors in Computing Systems}, CHI EA '17, pages 1624--1631, New York,
  NY, USA, 2017. ACM.

\bibitem{Costa:2016:ELB:2971648.2971752}
Jean Costa, Alexander~T. Adams, Malte~F. Jung, Fran\c{c}ois Guimbreti\`{e}re,
  and Tanzeem Choudhury.
\newblock Emotioncheck: Leveraging bodily signals and false feedback to
  regulate our emotions.
\newblock In {\em Proceedings of the 2016 ACM International Joint Conference on
  Pervasive and Ubiquitous Computing}, UbiComp '16, pages 758--769, New York,
  NY, USA, 2016. ACM.

\bibitem{Maekawa:2016:TPF:2971648.2971721}
Takuya Maekawa, Daisuke Nakai, Kazuya Ohara, and Yasuo Namioka.
\newblock Toward practical factory activity recognition: Unsupervised
  understanding of repetitive assembly work in a factory.
\newblock In {\em Proceedings of the 2016 ACM International Joint Conference on
  Pervasive and Ubiquitous Computing}, UbiComp '16, pages 1088--1099, New York,
  NY, USA, 2016. ACM.

\end{thebibliography}

\end{document}